\documentstyle[12pt]{article}

\newcommand {\vs}[1]  { \vspace*{#1 cm} }

\newcounter{eq}
\newcounter{sc}

\newcommand {\MPL}  {Mod. Phys. Lett.}
\newcommand {\IJMP}  {Int. J. Mod. Phys.}
\newcommand {\NP}   {Nucl. Phys.}

\newcommand {\PL}   {Phys. Lett.}
\newcommand {\PR}   {Phys. Rev.}

\def\overleftrightarrow#1{\vbox{\ialign{##\crcr
 $\leftrightarrow$\crcr\noalign{\kern-1pt\nointerlineskip}
 $\hfil\displaystyle{#1}\hfil$\crcr}}}

\newlength{\minitwocolumn}
\begin{document}

\begin{flushright}
DPUR/TH/6\\
September, 2007\\
\end{flushright}
\vspace{30pt}
\pagestyle{empty}
\baselineskip15pt

\begin{center}
{\large\bf Gravitational Higgs Mechanism with a Topological Term

 \vskip 1mm
}

\vspace{20mm}

Ichiro Oda
          \footnote{
           E-mail address:\ ioda@phys.u-ryukyu.ac.jp
                  }
\\
\vspace{10mm}
          Department of Physics, Faculty of Science, University of the 
           Ryukyus,\\
           Nishihara, Okinawa 903-0213, JAPAN \\

\end{center}


\vspace{20mm}
\begin{abstract}
We investigate the Higgs mechanism for gravity, which has been recently put forward by
't Hooft, when the Polyakov-type action for scalar fields is added to the original action.  
We find that from the Polyakov-type action, it is very natural to derive an 'alternative' 
metric tensor composed of the scalar fields. The positivity condition on the determinant can 
be also derived easily by requiring that this term does not change the dynamics at all
and becomes a topological number, that is, the wrapping number. It turns out that the gauge conditions 
adopted by 't Hooft are nothing but the restriction on a sector with unit wrapping number.

\vspace{15mm}

\end{abstract}

\newpage
\pagestyle{plain}
\pagenumbering{arabic}


\rm
\section{Introduction}

There have been recently some interesting works about the spontaneous 
symmetry breakdown (SSB) of general coordinate reparametrization invariance
\cite{Kaku1, Porrati, Kirsch, 't Hooft}.
The first motivation behind these works comes from brane world scenario 
where the presence of a brane breaks some of diffeomorphisms in the directions 
perpendicular to the brane spontaneously, 
so that we expect that there might naturally appear a gravitational 
Higgs mechanism in this context \cite{Kaku1, Porrati}. 

The second motivation is that this phenomenon might play
an important role in developing string theory approach to quantum 
chromodynamics (QCD) in future \cite{'t Hooft}.
For instance, if we wish to apply a bosonic string theory
to the gluonic sector in QCD, massless fields such as tachyonic 
scalar and spin 2 gravitons in string theory, must become massive or be 
removed somehow by some ingenious dynamical mechanism since such the fields 
do not exist in QCD.

As the final motivation, SSB of general coordinate reparametrization
invariance might lead to some resolution for cosmological problems such as 
cosmological constant problem \cite{Kirsch, 't Hooft}.

Recently, 't Hooft has proposed an interesting Higgs mechanism for gravity
where massless gravitons '$\it{eat}$' four real scalar fields, thereby 
becoming massive \cite{'t Hooft}. His motivation mainly lies in string theory 
approach to QCD and if the approach is effective, the massless gravitons must 
acquire a huge mass (perhaps, the Planck mass) and become unobservable at least
in the low energy region. In this model, diffeomorphisms are broken by four
real scalar fields spontanously such that vacuum expectation values (VEV's) of
the scalar fields are chosen to the four space-time coordinates up to a
proportional constant by gauge-fixing diffeomorphisms. Of course, the number 
of dynamical degrees of freedom is unchanged before and after SSB. Actually, 
before SSB of diffeomorphisms there are massless gravitons of two dynamical 
degrees of freedom and four real scalar fields whereas after SSB we have massive 
gravitons of five dynamical degrees of freedom and one real scalar field 
so that the number of dynamical degrees of freedom is equal to six 
both before and after SSB as desired. 

A key observation in the 't Hooft model is that the scalar field appearing after SSB is a 
non-unitary propagating field so that in order to avoid violation of unitarity
it must be removed from the physical Hibert space in terms of some 
procedure.\footnote{More recently, a model of gravitational 
Higgs mechanism without the non-unitary propagating scalar field was constructed 
in a conformally flat expanding background in Ref. \cite{Kaku2}.} In fact, two
methods were proposed at the classical level \cite{'t Hooft}. 
One method is to require that the energy-momentum tensor of the matter field 
does not couple to the usual metric tensor but the modified metric one 
in such a way that the non-unitary scalar field does not couple to 
the energy-momentum tensor directly. Another method is to eliminate the 
time-like component of the scalar fields by imposing a constraint on the
scalar fields.\footnote{In order to match the Minkowskian signature of the
background, one of the four scalar fields must take a negative signature.
Or equivalently, in the Euclidean signature after the Wick rotation, one
of the four scalar fields must have an imaginary vacuum expectation value.}
It is worthwhile to notice that in both the methods we have to
introduce an '$\it{alternative}$' metric tensor constructed out of four
real scalar fields, but its derivation from the first principle is 
lacking.\footnote{Here we would like to emphasize that there is no rule 
that we have only unique metric tensor in our world. For instance, there might be
a possibility such that we have two distinct metric tensors in our world, 
one of which controls the macroscopic, cosmological 
region while the other metric tensor does the microscopic, elementary particles' region.
Then, a real problem is to understand the relationship between two metric tensors
in the intermediate region.} 

In this short article, we investigate a possibility of having a topological
term in the 't Hooft theory. Topology has thus far played a central role 
in quantum field theories so it is worth pursuing such a possibility. 
To do that, we incorporate the Polyakov-type action to the 't Hooft's
starting action and explore how this term behaves when we require that 
this additional term should not change the dynamics completely. 
Similar but different approaches have been already taken into consideration 
in Ref. \cite{Oda1}.

Let us start with the following Euclidean action in four space-time dimensions. 
This action differs from the 't Hooft action \cite{'t Hooft} only by the last term $S_P$:
\begin{eqnarray}
S = S_{EH} + S_\Lambda + S_\phi + S_M + S_P,
\label{1.1}
\end{eqnarray}
where each term takes the following form:
\begin{eqnarray}
S_{EH} &=& \frac{1}{16 \pi G} \int d^4 x \sqrt{g} R, \nonumber\\
S_\Lambda &=& - \frac{\Lambda}{8 \pi G} \int d^4 x \sqrt{g}, \nonumber\\
S_\phi &=& - \frac{1}{2} \int d^4 x \sqrt{g} g^{\mu\nu}
\partial_\mu \phi^a \partial_\nu \phi^a, \nonumber\\
S_M &=& \int d^4 x {\cal{L}}_{matters}, \nonumber\\
S_P &=& - \frac{T}{2} \int d^4 x \sqrt{g_\phi} g^{\mu\nu}_\phi
\partial_\mu \phi^a \partial_\nu \phi^a
+ \Lambda_P \int d^4 x \sqrt{g_\phi}.
\label{1.2}
\end{eqnarray}
Here the fourth term $S_M$ describes an action for a general matter field but $\phi^a$.

Now let us take a variation with respect to $g^{\mu\nu}_\phi$,
which gives us the famous equations of motion in string (or brane) theory:
\begin{eqnarray}
0 &=& T_{\mu\nu}^\phi   \nonumber\\
&=& \partial_\mu \phi^a \partial_\nu \phi^a 
- \frac{1}{2} g_{\mu\nu}^\phi g^{\gamma\delta}_\phi 
\partial_\gamma \phi^a \partial_\delta \phi^a 
+ \frac{\Lambda_P}{T}  g_{\mu\nu}^\phi.
\label{1.3}
\end{eqnarray}
It is easy to solve the equations whose result reads
\begin{eqnarray}
g_{\mu\nu}^\phi = \frac{T}{\Lambda_P} \partial_\mu \phi^a \partial_\nu \phi^a. 
\label{1.4}
\end{eqnarray}
{}For simplicity, we henceforth assume $T = \Lambda_P$.\footnote{A different value
from this value is in essence equivalent, if we make a suitable rescaling
of $g_{\mu\nu}^\phi$. Then, this results in a multicative coefficient 
in the relation between $g_{\mu\nu}^\phi$ and $\partial_\mu \phi^a \partial_\nu \phi^a$.} 
In this way, we arrive at the expression of an '$\it{alternative}$' metric tensor 
constructed out of four real scalar fields 
\begin{eqnarray}
g_{\mu\nu}^\phi = \partial_\mu \phi^a \partial_\nu \phi^a.
\label{1.5}
\end{eqnarray}
Note that this relation was assumed in an ad hoc manner in Ref. \cite{'t Hooft}
whereas it is now derived from the action principle.

Next, let us rewrite the Polyakov-type action to the Nambu-Goto-type one
by substituting the relation (\ref{1.5}) into $S_P$:
\begin{eqnarray}
S_P &=& - \Lambda_P \int d^4 x \sqrt{g_\phi}  \nonumber\\
&=& - \Lambda_P \int d^4 x 
\sqrt{\det_{\mu, \nu} \partial_\mu \phi^a \partial_\nu \phi^a},
\label{1.6}
\end{eqnarray}
where we have appended the indices $\mu, \nu$ to the determinant
in order to emphasize that this is the determinant for a matrix
with row index $\mu$ and column one $\nu$. Now one finds that 
one can recast this equation further when the number of space-time
dimensions is equal to that of scalar fields, which just corresponds
to the situation at hand:
\begin{eqnarray}
S_P = - \Lambda_P \int d^4 x 
| \det_{\mu, a} \partial_\mu \phi^a |,
\label{1.7}
\end{eqnarray}
where the determinant with respect to $\mu$ and $\nu$ is replaced with
the one with respect to $\mu$ and $a$. Moreover, we should take the
absolute value of the determinant.\footnote{Some references miss
putting the absolute value.}
 
Note that at this stage the term $S_P$ is almost topological in the
sense that at least locally one can eliminate all the dynamical degrees 
of freedom associated with four scalar fields by using diffeomorphisms 
in four space-time dimensions. Nevertheless this term is not completely 
topological in that there is an ambiguity coming from the absolute value.  
Then, we require that this additional term $S_P$ should not change 
the dynamics of the original 't Hooft theory both locally and globally.
In order to do so, we must pick up either a positive sign or a negative one.
From now on, we shall confine ourselves to a positive sign:
\begin{eqnarray}
\det_{\mu, a} \partial_\mu \phi^a > 0.
\label{1.8}
\end{eqnarray}
Then, we have a completely topological term for $S_P$:
\begin{eqnarray}
S_P &=& - \Lambda_P \int d^4 x \det_{\mu, a} \partial_\mu \phi^a  \nonumber\\
&=& - \Lambda_P \int d^4 x \varepsilon^{\mu\nu\rho\sigma} 
\partial_\mu \phi^1 \partial_\nu \phi^2 \partial_\rho \phi^3 \partial_\sigma \phi^4,
\label{1.9}
\end{eqnarray}
which is nothing but the wrapping number $\Pi_3(S^3) = Z$ up to
an overall constant.\footnote{In case of the 't Hooft theory, the four-dimensional
space-time has an asymptotically flat boundary and after topological 
compactification with one point added, the boundary becomes $S^3$.}

Here let us consider the gauge conditions for diffeomorphisms in \cite{'t Hooft}, 
which are given by
\begin{eqnarray}
\phi^\mu = m x^\mu.
\label{1.10}
\end{eqnarray}
It was pointed out in \cite{'t Hooft} that there are ambiguities
in the gauge-fixing conditions (\ref{1.10}), for which the
condition (\ref{1.8}) is imposed by hand. Recall that
this condition also emerges in the cure of the indefinite metric problem.
From the present point of view, the condition (\ref{1.8}) appears
in order to make the Polyakov-type term completely topological.
Then, what is the mathematical meaning of the gauge-fixing conditions (\ref{1.10})?
As is easily shown, with the gauge-fixing conditions (\ref{1.10}) 
the wrapping number takes one (if we take $m=1$), 
so the gauge conditions (\ref{1.10}) mean that we are in a topological 
sector with unit wrapping number. To put differently, together with
the condition (\ref{1.8}) and the gauge-fixing conditions (\ref{1.10}),
the 't Hooft theory is uniquely defined in the Hilbert space with
unit wrapping number. In this context, one can conjecture that the
model might be generalized to the more general Hilbert space
where the wrapping number takes a more general value, by choosing
different conditions from (\ref{1.8}) and (\ref{1.10}). 

In conclusion, in this article, we have investigated a possibility
of having a topological term within the framework of the Higgs
mechanism for gravity. The results obtained so far make clear
that such a topological approach is very useful even in this theory
as in the conventional quantum field theories. We have derived
an '$\it{alternative}$' metric tensor constructed out of four
real scalar fields by starting with the action. Furthermore,
we have shed a new light on the interpretation
for both the positivity of the determinant and the gauge-fixing 
conditions.

Finally, we wish to comment on two methods of removing a non-unitary 
propagating scalar field, which were presented in \cite{'t Hooft}.  
In these methods, we have to impose an additional constraint by hand.
Although such a procedure is reasonable at the classical level,
it might lead to some inconsistency at the quantum level. Maybe, a more
plausible method is to introduce such a constraint in the theory
as the gauge-fixing condition for an extra local symmetry.
As one of such the methods, in future we would like to take account of
a theory where there is the gauge field $A_\mu$ and one real scalar
since the total number of dynamical degrees of freedom is three, which
is the minimum number for the gravitational Higgs mechanism in four
dimensions \cite{Oda2}. In this method, the time-like component $A^0$ 
in the gauge field could be removed through the usual gauge invariance.

\begin{flushleft}
{\bf Note added}
\end{flushleft}

During preparation of this article, a preprint \cite{Kaku2} has
appeared where an idea of introducing the gauge field and one real
scalar was commented. Moreover, more recently, a related preprint \cite{Jackiw}
has been also put on the archive.\footnote{We wish to thank D. Sorokin
for informing me of the preprint \cite{Jackiw}.}

\begin{flushleft}
{\bf Acknowledgement}
\end{flushleft}

We would like to thank D. Sorokin and M. Tonin for valuable 
discussions. Part of this work has been done during stay
at Dipartimento di Fisica, Universita degli Studi di Padova.
We also wish to thank for a kind hospitality.

\vs 1   

\end{document}